\documentclass[conference]{IEEEtran}
\IEEEoverridecommandlockouts
\def\BibTeX{{\rm B\kern-.05em{\sc i\kern-.025em b}\kern-.08em
    T\kern-.1667em\lower.7ex\hbox{E}\kern-.125emX}}
\usepackage{mathtools}

\usepackage[nodisplayskipstretch]{setspace}
\usepackage[utf8x]{inputenc}
\usepackage{tikz}

\newcommand\tx{\tilde{X}(t)} 
\newcommand\hx{\hat{X}(t)}
\newcommand\ro{\rho(t)}
\newcommand\cs{c_{\mathrm{s}}}
\newcommand\ct{c_{\mathrm{t}}}

\newcommand\brq{\bar{q}} 
\newcommand\brp{\bar{p}} 
\newcommand\brmu{\bar{\mu}}
\DeclareMathOperator*{\argmin}{arg\,min} 
\usepackage{subfigure}
\usepackage{moreverb}
\usepackage{epsfig}
\usepackage{amsmath,amssymb,amsthm,mathrsfs,amsfonts,dsfont}
\usepackage{tikz}
\usepackage{fancyhdr}
\usepackage{adjustbox,lipsum}
\usepackage[font={small}]{caption}
\usepackage{color,soul}
\usepackage{amsmath}
\usepackage{amsthm,amssymb,amsmath,bm}
\usepackage{subfigure}
\usepackage{amsfonts}
\usepackage{amsmath}
\usepackage{epsfig}
\usepackage{amssymb}
\usepackage{amsmath}
\usepackage{cite}
\usepackage{bbm}
\newtheorem{Pro}{Proposition}

\begin{document}
\title{ \huge 
 Optimal Semantic-aware Sampling and Transmission   in Energy Harvesting Systems Through the AoII

}

 \author{\IEEEauthorblockN{
 Abolfazl Zakeri\IEEEauthorrefmark{1}, Mohammad Moltafet\IEEEauthorrefmark{2},
 and Marian Codreanu\IEEEauthorrefmark{3}}
 \vspace{-1 em }
 \\
 \IEEEauthorrefmark{1}\normalsize Centre for Wireless Communications – Radio Technologies,
University of Oulu, Finland,
 email: 
abolfazl.zakeri@oulu.fi
\\
\IEEEauthorrefmark{2}Department of Electrical and Computer Engineering, University of California Santa Cruz,
email:
mmoltafe@ucsc.edu
\\
\IEEEauthorrefmark{3}Department of Science and Technology,
 Link\"{o}ping University, Sweden, 
 email: marian.codreanu@liu.se
\thanks{This research has been financially supported by the Academy of Finland (grant 323698), and the 6G Flagship program (grant 346208).  The work of M. Codreanu has also been financially supported in part by the Swedish Research Council (grant 2022-03664). We would like to gratefully acknowledge the contributions and insights of Markus Leinonen to this paper.}
}

\maketitle

	\begin{abstract}
 We study a real-time tracking problem in an energy harvesting status update system with a Markov source and an imperfect channel, considering both sampling and transmission costs. 
 The problem's primary challenge stems from the non-observability of the source due to the sampling cost. 
By using the age of incorrect information (AoII) as a semantic-aware performance metric, our main goal is to find an optimal policy  that minimizes the time average AoII subject to an energy-causality constraint.
To this end, a stochastic optimization problem is formulated and solved by modeling it as a partially observable Markov decision process (POMDP). 
More specifically, to solve the main problem, we use the notion of  a belief state and  cast the problem as a belief MDP problem.
Then, for the perfect channel setup,  we effectively truncate the corresponding belief space and solve the MDP problem using the relative value iteration method.
For the general setup, a deep reinforcement learning policy is proposed.
The simulation results show the efficacy of the derived policies in comparison to an AoI-optimal policy and an opportunistic baseline policy.
\end{abstract}
\section{ Introduction} 
The age of information (AoI) has been  introduced to quantify the  information freshness in status update systems \cite{Roy_2012}. 
Since then, there has been significant research on the AoI in different areas, e.g., queuing systems \cite{Marian_Information}, and  scheduling and sampling problems \cite{Zakeri_Journal_Relay, Eytan_Sch2}. 
Besides the AoI, various related metrics have been also proposed, e.g.,  the value of information \cite{Nikol_VoI}
and the age of incorrect information (AoII) \cite{Tony_1}, which also accounts for the value of information. 
The AoII essentially amalgamates time penalty with accuracy/distortion penalty
 to quantize the discrepancy between the information source and its estimation on the monitor side, 
capable of capturing the semantics/meanings of data transfer \cite{Deniz_Semantic_JSAC,ELIF_Semantic_mag, AoII_Semantic}, the provisioning of the right  piece of information to the right point of computation (or actuation) at the right
point in time \cite{ELIF_Semantic_mag}.

The ultimate goal in status update systems is the real-time tracking  of a real-world stochastic process on the side of a remote monitor \cite{Kam_Towards_eff_2018}. Recently, the real-time tracking problem has been studied in a handful of papers, e.g.,  the papers \cite{Yin_Sun_IT, Atilla_RtM} which used distortion-based metrics, the work in \cite{Nikolaos_goal_or} which used goal-oriented metrics, and the works \cite{Kam_2018,Assaad_AoII_Not_obs, 
Assaad_unknownSource, Petar_AoII_ICC,Chen_AoII_2023} which used the AoII metric.
In this paper, we study the real-time tracking problem with the AoII metric. 

The work \cite{Kam_2018} provided the AoI-, the real-time error-, and the  AoII- optimal policies for the remote tracking problem of a symmetric binary Markov source in a source-monitor-paired system. Their results show that the {sample-at-change} policy, which simultaneously samples and transmits  whenever there is a difference (or a change) between the source state and its estimation, optimizes both the real-time error and the AoII. 
The authors of \cite{Assaad_AoII_Not_obs} studied the remote tracking problem of a Markov source in a multi-source setup, where the decision-maker resides on the monitor side. 
They developed a heuristic scheduling policy that minimizes the mean AoII using the partially observable Markov decision process (POMDP) framework and the idea of the Whittle index policy. They then 
optimized the AoII under an unknown Markov source in \cite{Assaad_unknownSource}. 

However,  in most of  the works on the AoII, the AoII optimization relies on \textit{fully observable} source, e.g.,  \cite{Tony_1, AoII_Semantic, Kam_Towards_eff_2018, AoII_ARQ, Petar_AoII_ICC, Kam_2018, Chen_AoII_2023}.
Having a fully observable source requires
continuous sampling and processing of the  source signal.
However, in practice, this could be challenging
due to high
sampling costs, or even impossible due to, e.g., insufficient energy to make sampling at each time, as is often the case in
energy harvesting systems.

To the extent of our knowledge, only \cite{Assaad_AoII_Not_obs}  (and its  subsequent extension \cite{Assaad_unknownSource}) optimized the AoII under a partially observable source where the  partial observability comes from the controller's location, whereas in this paper it is due to the sampling cost.
Furthermore, in \cite{Assaad_AoII_Not_obs}, the sampling of the source state and its immediate transmission are upon request at \text{any given time}. In contrast, in this paper,  both sampling and transmission operations are subjected to the stochastic availability of energy and are associated with distinct costs. Additionally, different to \cite{Assaad_AoII_Not_obs}, in our system, there is a possibility to retransmit an old sample.

The goal of this paper is to find an AoII-optimal \textit{joint} sampling and transmission policy in an energy-harvesting real-time tracking system under both sampling and transmission costs, where the  sampling cost renders the source unobservable. 
 We consider a discrete-time status update system with an imperfect channel consisting of a source, a sampler, a buffer-aided transmitter,  and a monitor, depicted in Fig.~\ref{Fig_EnrHar}.
 Operation of each sampling and transmission consumes some units of energy, imposing an energy causality constraint. 
 

We formulate a stochastic optimization problem aiming at minimizing the average AoII subject to the energy causality constraint. The problem is modeled as a POMDP that is subsequently turned into a belief MDP
problem. 
Solving the belief MDP is challenging owing to its continuous belief state space. Despite the challenge, for a perfect channel case, we effectively truncate the belief space and find an optimal policy via the relative value iteration (RVI) algorithm. Moreover, for the general imperfect channel case, a deep Q-network (DQN) policy is proposed. 
Simulation results are conducted to show the effectiveness of the derived policies compared to an AoI-optimal policy and an opportunistic baseline policy.

The rest of this paper is organized as follows. The system model and problem formulation are presented in Section \ref{Sec_SM_PF}.  The solution method is provided in Section \ref{Sec_Solution}. 
Finally,  the numerical results and conclusions are presented, respectively,  in Section \ref{Sec_NR} and Section \ref{Sec_Con}.
\section{System Model and Problem Formulation}\label{Sec_SM_PF}
\subsection{System Model}
\begin{figure}[t!]
    \centering
    \includegraphics[width=.5\textwidth]{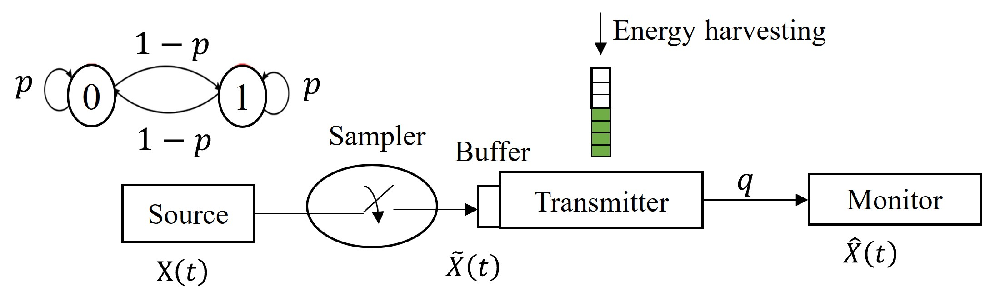}    
    \caption{System model.
    }
    \label{Fig_EnrHar}
\end{figure}

We consider a  real-time tracking system consisting of an information source, a sampler, a transmitter, and a monitor, as shown in Fig. \ref{Fig_EnrHar}. 
The system
is powered  by an energy-harvesting module equipped with a finite-capacity battery.
The monitor is interested in the 
 real-time tracking of the  source.
 To this end,
 the controller, located at the transmitter side,  should appropriately decide when to sample and when to transmit the taken sample.
Most importantly, the controller \textit{does not observe} the source unless a sample is
taken; the controller
 observes the battery level, the information in the transmitter's buffer, and the transmission results (i.e., ACK/NACK feedback from the monitor).

We assume a discrete-time system with unit time slots ${ t\in\{0,1,2,\ldots\} }$. 
The source is modeled via a  two-state (binary) symmetric  discrete-time Markov  process ${ X(t)\in\{0,1\} }$,   with the self-transition
probability $p$. 
Without loss of generality, we assume $p>0.5$ for the sake of presentation clarity. Note that the results are identical for $p<0.5$ provided that the monitor employs an appropriate state estimation strategy specified below.
The binary source is a commonly used model (e.g., \cite{Kam_2018, Kam_Towards_eff_2018, Tony_1, Chen_AoII_2023}) and it provides fundamental insights into the sampling and transmission optimization in the system.\footnote{An extension to a multi-state Markov source is deferred for future work.} 
The process $X(t)$  is \textit{not observable} at slot $t$ unless a  sample is taken.
Thus,  sampling at slot $t$ reveals  $X(t)$.

When requested by the controller, the sampling of the source takes place at the beginning of the slot, right after the state transition (it is assumed that  the source and the system clocks are synchronized).
The last taken sample  is always stored in the transmitter's buffer.
We denote the last sampled state of the source in the buffer at slot $t$ by $\tx$. 

\textit{Estimation strategy:}
The monitor needs to have real-time estimate of the source. We denote the source estimate at slot $t$ by $\hx$.
We assume that the monitor employs a maximum likelihood estimation, which for the source with  $p>0.5$ is the last received sample, and for $p<0.5$ is alternating the last received sample at each slot \cite{Kam_Towards_eff_2018}.

\textit{Communication channel:} We assume an imperfect channel between the transmitter and the monitor. Each transmission takes one slot
 and it is successfully received with probability ${ q} $, referred to as the reception success probability.
The  unsuccessfully received   samples  can be retransmitted, and they experience  the same reception  success probability.
We assume that  perfect  (i.e., instantaneous  and error-free) feedback is available for each transmission. 

\textit{Sampling and Transmission Costs:} We assume that
each sampling consumes $\cs$ units of energy (i.e., sampling cost), and
each transmission consumes $\ct$ units of energy (i.e., transmission  cost). The costs are assumed to be constant.

\textit{Decision/optimization variables:}
In each slot, the controller decides the sampling and the transmission decisions. 
Let $\alpha(t)\in\{0,1\}$ denote the transmission decision at slot $t$, where $\alpha(t)=1$ means transmitting a sample; otherwise, $\alpha(t)=0$.
Let $\beta(t)\in\{0,1\}$ denote the sampling decision at slot $t$, where $\beta(t)=1$ means sampling (and observing the source's current state); otherwise, $\beta(t)=0$.
We assume that in the case of the concurrent sampling and transmission, i.e., $\beta(t)=\alpha(t)=1$,  the transmitted sample is the current (updated) source state, i.e., $\tx =X(t)$.

\textit{Energy harvesting model:} The energy supplier of the system harvests energy and stores it in a finite-capacity battery of $E$ units of energy.
Similarly to, e.g.,~\cite{EH_bern_conf, EH_Berno_Erikl}, we assume that  the energy arrivals $u(t)$ follows 
a Bernoulli process with parameter $\mu$, i.e., ${\Pr\{u(t)=1\} = \mu}$. 
The battery level at slot $t$, denoted by $e(t)\in\{0,\dots,E\}$, evolves as 
\begin{equation}\label{Eq_enrgydynamic}
\begin{array}{ll}
e(t+1) = \min \left\{e(t) + u(t+1) - \big( \beta(t) \cs +  \alpha(t)\ct \big), E \right \}.
\end{array}
\end{equation}

\textit{The Age of Incorrect Information:}
We adopt the AoII used in \cite{Assaad_AoII_Not_obs} (and further studied in \cite{Assaad_unknownSource}).
The AoII here is the time elapsed since the last time when the source state was the same as the current estimate at the monitor, $\hat{X}(t)$.
Formally, let ${ V(t) \triangleq \max \{t'\le t: X(t') = \hat{X}(t)\} }$.
The AoII at slot $t$,  denoted by $\delta(t)$, is defined by  
\begin{align}\label{Eq_AoII_def}
     \delta(t) = \left(t - V(t) \right).
\end{align}

\subsection{Problem Formulation}
Given the above definitions, our goal is to solve
the following stochastic optimization problem:
        \begin{subequations}
       \label{Org_P1}
       \begin{align}
          {\mbox{minimize}}~~~   &
          \limsup_{T\rightarrow \infty}\,\frac{1}{T}   \sum_{t=1}^T \Bbb{E}\{ \delta(t) \}
        		\\
        		\mbox{subject to}~~~ & 
                   \label{Cons_smp_trans_S2}
                 e(t) - \beta(t) \cs-\alpha(t) \ct\ge 0, ~\forall\,t,
                   \end{align}
        		\end{subequations} 
          with variables ${ \{\alpha(t),\beta(t)\}_{t=1,2,\ldots} }$, 
where the constraint \eqref{Cons_smp_trans_S2} is the energy causality constraint.
Furthermore, $\mathbb{E}\{\cdot\}$ is the expectation notation which is taken with respect to the system's randomness (due to the source, the energy arrivals,  and the wireless channel) and the (possibly randomized) decision variables $\alpha(t)$ and $\beta(t)$ made in reaction to the available observations at the controller.
\section{ An Optimal Policy  }\label{Sec_Solution}
Here we  present an optimal policy for   problem~\eqref{Org_P1}.
Note that   AoII is a function of the source  $X(t)$ which is not observable due to the sampling cost. Thus, we first model problem \eqref{Org_P1} as a POMDP 
  and subsequently cast  it into an MDP problem. 

The POMDP is described by the following elements:
\\
$\bullet$
\textit{State:}
 Let $\ro$ be a binary indicator indicating whether the last sample at the buffer $\tx$ equals to the estimate $\hx$. 
Specifically, $\ro$ is defined as
   \begin{equation}
     \begin{array}{cc}
 \ro \triangleq
\left\{\begin{array}{ll}
 0, & \text{if}~~   \tilde{X}(t) = \hx,
   \\
1 , & \text{if}~~   \tilde{X}(t) \neq \hx.
 \end{array}\right. 
    \end{array}
 \end{equation}  
We define the state at slot $t$ by $s(t) = \left( e(t), \delta(t), \ro\right)$.
The state space is denoted by $\mathcal{S}$. 
\\
$\bullet$ \textit{Observation:}
The observation at slot $t$, denoted by $o(t)$,  is~$o(t) =  (e(t), \ro)$.
\\
$\bullet$
\textit{Action:} 
There are totally four possible actions at each slot. 
However, by taking into account  
the  goal of problem \eqref{Org_P1} (i.e., minimizing the average AoII): 
i) re-transmitting an old 
sample when  $\tx = \hx$,
or ii) transmitting a fresh 
sample when $X(t)=\hx$ are both only wasting energy without reducing the distortion; thus, the decision to simultaneously sample and transmit, i.e., $\alpha(t)=\beta(t)=1$,
can be encoded to the action of taking a sample and transmitting that sample only if $X(t)\neq \hx$,
and
ii) the decision to take
a new sample without  simultaneously transmitting it (i.e., $\beta(t)=1$ and $\alpha(t)=0$) can be eliminated without losing the optimality; this is  because, 
in the case where the sample will not be transmitted at a later time, it simply consumes energy without enhancing performance, and 
in the case where the sample is to be transmitted later, taking a fresh sample just before transmission instead always has a higher probability to rectify the estimate and hence improve performance.
Thus, the action space of the POMDP, shown by $\mathcal{A}$, has three elements which are specified in the following.
The action at slot
$t$ is defined by ${ a(t) \in\{0,1,2\} }$, where $a(t) = 0$ indicates that the sampler and transmitter stay idle, 
$a(t) = 1$ indicates  that the transmitter re-transmits the  sample in the buffer 
 (the action could be $a(t) = 1$ if there is enough energy for at least one transmission and  $\tx \neq \hx$),
and $a(t) = 2$ indicates that  the sampler takes a new sample and the transmitter transmits that sample when $X(t) \neq \hx$ (the action could be $a(t) = 2$ if there is enough energy for at least one sampling and one transmission). Actions are determined by a policy, denoted by $\pi$, which is a (possibly randomized) mapping from $\mathcal{S}$ to $\mathcal{A}$.
\\
$\bullet$
\textit{State Transition Probabilities:} 
The  transition probabilities from  current state ${s=(e,\delta,\rho)}$ to  next state ${s'=(e',\delta',\rho')}$ under a given action $a$ is denoted by  
$
{
        \Pr\{s'\,|\,s, a\} }. 
  $  
  To facilitate a compact description of $\Pr\{s'\,|\,s,a\}$, we employ the shorthand notations
${\brq \triangleq 1-q}$, 
${\brmu \triangleq 1-\mu}$,
${\brp \triangleq 1-p}$,
and 
${c \triangleq \cs+ \ct }$.
  Since for a given action and state, the evolution of $\rho$, the AoII, and the energy
arrival process are independent,
  the transition probabilities 
 can be written as $ {
 \Pr\{s'\,|\,s, a\} = 
 \Pr\{ \rho'\,\big|\,\rho,\,\delta,\, a\}
 \Pr\{\delta'\,|\,\delta,\,a\}  
 \Pr\{e'\,|\,e,\delta,\,a\} } $, where 
\begin{equation}
     \begin{array}{cc}
        \Pr\{ \rho'\,\big|\,\rho,\,\delta,\, a\}=\left\{ 
  \begin{array}{ll}
  1,   &  \text{if} ~ a = 0, ~\rho' = \rho, 
  \\
  q,   &  \text{if} ~ a = 1, ~\rho' = 0,
  \\
 \brq,   &  \text{if} ~ a = 1, ~\rho' = \rho,
 \\
  1,   &  \text{if} ~ a = 2,~\delta = 0, ~\rho' = 0,
  \\
  q,   &  \text{if} ~ a = 2,~\delta \neq 0, ~\rho' = 0,
  \\
 \brq,   &  \text{if} ~ a = 2,~\delta \neq 0, ~\rho' = 1,
 \\
 0, & \text{otherwise}.
    \end{array}
    \right.
     \end{array}
 \end{equation}
\begin{equation}
     \begin{array}{ll}
        \Pr\{\delta'\,\big|\,\delta,\,a=0\}=\left\{  
  \begin{array}{ll}
  p,   &  \text{if} ~ \delta = 0,~\delta' = 0,  
  \\
  \brp,   &  \text{if} ~ \delta = 0,~\delta' = 1,
  \\
\brp,   &  \text{if} ~ \delta \neq 0,~\delta' = 0, 
 \\
p,   &  \text{if} ~ \delta \neq 0,~\delta' = \delta + 1,
 \\
 0, & \text{otherwise}.
    \end{array}
    \right.
     \end{array}
 \end{equation}
 \begin{equation}
     \begin{array}{ll}
        \Pr\{\delta'\,\big|\,\delta,\,a =  1\}=\left\{ 
  \begin{array}{ll}
  q \brp,   &  \text{if} ~ \delta = 0,~\delta' = 0,  
  \\
  qp,   &  \text{if} ~ \delta = 0,~\delta' = 1,  
  \\
  \brq p,   &  \text{if} ~ \delta = 0,~\delta' = 0,  
  \\
 \brq \brp,   &  \text{if} ~ \delta = 0,~\delta' = 1, 
  \\
 qp,   &  \text{if} ~ \delta \neq 0,~\delta' = 0,
  \\
 q \brp,   &  \text{if} ~ \delta \neq 0,~\delta' = \delta+1, 
 \\
 \brq \brp,   &  \text{if} ~ \delta \neq 0,~\delta' = 0,
\\
\brq p,   &  \text{if} ~ \delta \neq 0,~\delta' = \delta+1,
 \\
 0, & \text{otherwise}.
    \end{array}
    \right.
     \end{array}
 \end{equation}
 \begin{equation}
     \begin{array}{ll}
        \Pr\{\delta'\,\big|\,\delta,\,a =  2\}=\left\{ 
  \begin{array}{ll}
  p,   &  \text{if} ~ \delta = 0,~\delta' = 0,  
  \\
 \brp,   &  \text{if} ~ \delta = 0,~\delta' = 1,  
  \\
 qp,   &  \text{if} ~ \delta \neq 0,~\delta' = 0,
  \\
 q \brp,   &  \text{if} ~ \delta \neq 0,~\delta' = 1, 
 \\
\brq \brp,   &  \text{if} ~ \delta \neq 0,~\delta' = 0,
\\
\brq p,   &  \text{if} ~ \delta \neq 0,~\delta' = \delta+1,
 \\
 0, & \text{otherwise},
    \end{array}
    \right.
     \end{array}
 \end{equation} 
 \begin{equation}\label{Eq_Tp_dis_e}
     \begin{array}{ll}
        & \Pr\{e'\,\big|\,e,\delta,\,a\}   =
        \\&
        \left\{ 
  \begin{array}{ll}
  \mu,   & \text{if} ~ a = 0,~ e' = \min\{e+1, E\},
  \\
   \brmu,   & \text{if} ~ a = 0,~ e' = e,
    \\
     \mu,   & \text{if} ~ a = 1,~ e' = e+1 - \ct,
  \\
    \brmu,   & \text{if} ~ a = 1,~ e' = e - \ct,
  \\
  \mu,   & \text{if} ~ a = 2,~ e' = e+1-c,~\delta \neq 0,
  \\
  \brmu,   & \text{if} ~ a = 2,~ e' = e-c,~\delta \neq 0,
    \\
  \mu,   & \text{if} ~ a = 2,~ e' = e+1-\cs,~\delta = 0,
  \\
\brmu,   & \text{if} ~ a = 2,~ e' = e-\cs,~\delta = 0,
  \\
     0,  & \text{otherwise}.
    \end{array}
    \right.
     \end{array}
 \end{equation}
\\
$\bullet$
\textit{Observation function:}
The observation function is ${ \Pr\{o(t)\,|\,s(t),a(t-1)\} }$, which  is a deterministic function, i.e., ${ \Pr\{o(t)\,\big|\,s(t),a(t-1)\}  =  \mathds{1}_{\{o(t)=\left(e(t),  \delta(t), \rho(t)\right)\}} }$.
 \\
 $\bullet$
 \textit{Cost function:}
 The immediate cost function at slot $t$ is defined by $C(s(t))= \delta(t)$. 

\textit{Belief MDP Formulation:} To have optimal decision-making,  we
need to define state-like quantities that preserve the Markov
property and summarize all the necessary information called sufficient information states. 
Widely used sufficient states, as in this paper, are \textit{belief states} \cite[Ch. 7]{POMDP_AI}. 

Let  $I_{\mathrm{C}}(t)$ denote the complete information state at slot $t$ consisting of \cite[Ch. 7]{POMDP_AI}:
 i)     the initial probability distribution over states,
ii)  all past and current observations, i.e., 
${(o(0), \dots , o(t))}$,
and iii) all past actions, i.e., ${(a(0), \dots, a(t-1) )}$.
We define a belief  $b_i(t)$ by
\begin{equation}
     b_i(t)\triangleq \Pr\left\{\delta(t) = i\,\big|\,I_{\mathrm{C}}(t)\right\},~\, 
     i = 0,1,\dots,
 \end{equation}
The belief is updated
 as a function of current belief $\{b_i(t)\}_{i= 0,1,\dots}$,
the observation $o(t+1)$,
and current action $a(t)$. The following proposition gives the belief update. 
 \begin{Pro}\label{Prop_AoII_blf_gnr}
   Given  belief $\{b_i(t)\}_{i= 0,1,\dots}$,
 observation ${o(t+1)}$,
and  action $a(t)$, the belief update function is given by the following equations:
    \\
If $a(t) = 0$, \text{or} $a(t) = 1, \rho(t+1)=1$:
  \begin{equation}\label{Eq_BlfupdAoII_idle}
  \hspace{-1 em}
     \begin{array}{ll}
 b_i(t+1) =
 \left\{ 
  \begin{array}{ll}
   b_0(t)p + \left(1-b_0(t) \right) (1-p), & i=0,
   \\
 (1-p)  b_{0}(t)  , & i= 1,
    \\
 p  b_{i-1}(t)  , & i= 2,3,\ldots, 
\end{array}\right. 
    \end{array}
 \end{equation}
 if $a(t) =1,  \rho(t+1) =0$:
 \begin{equation}
 \hspace{-1 em}
     \begin{array}{ll}
 b_i(t+1) =
 \left\{ 
  \begin{array}{ll}
 b_0(t)(1-p) + (1-b_0(t))p, & i=0,
   \\
  b_{0}(t) p  , & i= 1,
    \\
 (1-p)  b_{i-1}(t)  , & i= 2,3,\ldots, 
\end{array}\right. 
    \end{array}
 \end{equation}
  if $a(t) = 2, \rho(t+1)=1$:
  \begin{equation}
     \begin{array}{ll}
 b_i(t+1) =
 \left\{ 
  \begin{array}{ll}
  1-p, & i=0,
    \\
b_{i-1}(t) p  , & i= 1, 2,\ldots, 
\end{array}\right. 
    \end{array}
 \end{equation}
 and if $a(t) = 2, \rho(t+1)=0$:
  \begin{equation}\label{Eq_BlfupAoII_reset}
     \begin{array}{ll}
 b_i(t+1) =
 \left\{ 
  \begin{array}{ll}
  p, & i=0,
   \\
 1-p,   & i= 1,
    \\
0  , & i= 2,3,\ldots. 
\end{array}\right. 
    \end{array}
 \end{equation}
 \end{Pro}
Having the belief defined, we  formulate a belief
MDP by defining its state as
\begin{equation}
    \begin{array}{cc}
        z(t) \triangleq \left( e(t), \{b_i(t)\}_{i= 0,1,\dots},\ro \right),
    \end{array}
\end{equation}
and its immediate cost function as the expected AoII given by
$
{
C(z(t)) = \sum_{{i= 0,1,\ldots}} b_i(t)i.
}
$

Let $\mathcal{Z}$ denote the state space of the belief MDP, then, the goal is to find the optimal policy ${\pi^*: \mathcal{Z}\rightarrow \mathcal{A}}$ that is a solution to the following MDP problem:
\begin{align}\label{Eq_Prob_AoIIMDP}
\nonumber
   &    \pi^*({z(0)}) = 
       \\& 
       \argmin_{\pi\in\Pi}\left\{ \limsup_{T\rightarrow \infty} \frac{1}{T}  \sum_{t=1}^T \Bbb{E}\{ C(z(t))\,\big|\,z(0) \}
      \right\},
\end{align}
where the expectation is with respect to the policy and the system randomness, and $\Pi$ is the set of all admissible policies.

The state space of the belief MDP problem \eqref{Eq_Prob_AoIIMDP} is an infinite set, thus, finding an optimal policy is extremely challenging (see, e.g., \cite[Sec. 7.3]{POMDP_AI}); Actually, the problem is PSPACE-hard even for a finite horizon \cite[Sec. 7.3]{POMDP_AI}. 
Nonetheless, we will provide an optimal policy via the RVI algorithm
for the case where the channel is perfect, i.e., $q=1$,
and 
propose an online learning-based algorithm for the general  case. 
\subsubsection{An Optimal Policy Under The Perfect Channel}
It can be observable that, under the perfect channel, the re-transmission action $a(t)=1$ is unnecessary (as always $\tx=\hx$) so actions are essentially the idle action $a(t)=0$ and the sample and transmission action $a(t)=2$. 

Thus, the belief update  follows  \eqref{Eq_BlfupdAoII_idle} or \eqref{Eq_BlfupAoII_reset} depending on the actions taken. Next,   we will characterize and effectively truncate the belief space using the AoI at the transmitter $\theta(t)$, which allows us  to find an optimal policy.  

The following proposition shows one-to-one mapping between $\theta(t)$ and belief $\{b_i(t)\}_{i=0,1,\dots}$. 
\begin{Pro}\label{Prop_AoII_blf_prfch}
    Suppose $\theta(t) = n$, $n=1,2,\dots$. Then, for the perfect channel (i.e., $q=1$), the belief at slot $t$ is given by 
   \begin{equation}\label{Eq_AoII_BD}
     \begin{array}{cc}
 b_i(t) =
\left\{\begin{array}{ll}
  g(n), & i=0,
   \\
g(n-i)(1-p)p^{(i-1)} , & i= 1,\dots,n,
   \\
  0, & i=n+1,\dots,
 \end{array}\right. 
    \end{array}
 \end{equation}
 where the function $g(n)$ is given by
 \begin{equation}
\hspace{-1 em}
     \begin{array}{ll}
        g(n) \triangleq \Pr\{\delta(t) = 0\,\big|\,\theta(t)=n\}
 =
 0.5(1+(2p-1)^n),
     \end{array}
 \end{equation}
 and $g(0) \triangleq 1$.
\end{Pro}
One can observe from \eqref{Eq_AoII_BD}  that for sufficiently large values of  the AoI $\theta(t)$, denoted by $\bar{N}$,   the belief corresponding to $\theta(t)=\bar{N}$, converges to the following: 
    \begin{equation}
    \hspace{-1 em}
     \begin{array}{ll}
 b_i(t) =
\left\{\begin{array}{ll}
  0.5, & i=0,
   \\
g(\bar{N}-i)(1-p)p^{(i-1)} , & i= 1,\dots,\bar{N},
   \\
  0, & i=\bar{N}+1,\dots.
 \end{array}\right. 
    \end{array}
 \end{equation}
 Thus, we can effectively truncate the belief space by bounding the AoI with $\bar{N}$. 

We have shown that both the cost function and belief state of problem \eqref{Eq_Prob_AoIIMDP} can be written only as a function of the AoI $\theta(t)$, which is bounded by $\bar{N}$.
Thus, for the perfect channel, problem \eqref{Eq_Prob_AoIIMDP} can be expressed as a finite-state MDP problem with the following elements:
 \\
$\bullet$
\textit{State:}
The state at slot $t$ is $\underline{s}(t)= \left( e(t), \theta(t)\right),$ where $\theta(t)\in\{1,2,\dots,\bar{N}\}$. 
The state space is denoted by $\underline{\mathcal{S}}$, which
is a finite set.
\\
$\bullet$
\textit{Action:} 
The actions are $a(t)=0$ and $a(t)=2$.
\\
$\bullet$
\textit{State Transition Probabilities:} 
The  transition probabilities from  current state ${\underline{s} =(e,\theta)}$ to  next state ${\underline{s}'=(e',\theta')}$ under a given action $a$ is defined by  
$
{
        \Pr\{\underline{s}'\,|\,\underline{s}, a\} }, 
  $  
which can be written as $ {\Pr\{\theta'\,|\,\theta,\,a\}\Pr\{e'\,|\,e,\theta,a\} } $, where 
\begin{equation}\label{Eq_AoIDyn}
 \hspace{-0.5 em}
     \begin{array}{cc}
        \Pr\{\theta'\,|\,\theta,\,a\} =\left\{ 
  \begin{array}{ll}
  1,   & \text{if} ~ a \neq 2,~ \theta' = \min(\theta+1,\bar{N}),
  \\
     1,  & \text{if} ~ a = 2,~ \theta'=1,
     \\
     0,  & \text{otherwise},
    \end{array}
    \right.
     \end{array}
 \end{equation}
\begin{equation}
\hspace{-1 em}
     \begin{array}{ll}
     &    \Pr\{e'\,|\,e,\theta,\,a\}  =
        \\&
        \left\{ 
  \begin{array}{ll}
  \mu,   & \text{if} ~ a = 0,~ e' = \min\{e+1, E\},
  \\
    \brmu,   & \text{if} ~ a = 0,~ e' = e,
  \\
  \mu(1-g(\theta)),   & \text{if} ~ a = 1,~ e' = e+1-c,
  \\
  \brmu (1-g(\theta)),   & \text{if} ~ a = 1,~ e' = e-c,
    \\
  \mu g(\theta),   & \text{if} ~ a = 1,~ e' = e+1-\cs,
  \\
 \brmu g(\theta),   & \text{if} ~ a = 1,~ e' = e-\cs,
  \\
     0,  & \text{otherwise},
    \end{array}
    \right.
     \end{array}
 \end{equation}
 where  $g(\theta) = 0.5(1+(2p-1)^\theta)$.
 \\
 $\bullet$
 \textit{Cost Function:}
 The immediate cost function at slot $t$ is the expected AoII  given by
 \begin{equation}
 \begin{array}{cc}
 \textstyle
   C(\underline{s}(t)) = \sum_{i=0}^{\theta(t)} b_i(t)i,
      \end{array}
 \end{equation}
where $b_i(t)$ is given by \eqref{Eq_AoII_BD}.

Having the MDP specified above, we apply the RVI algorithm  to find an optimal policy for problem \eqref{Eq_Prob_AoIIMDP} under the perfect channel. 
The RVI algorithm transforms the Bellman's optimality equation into the following iterative process for each state $s\in\underline{\mathcal{S}}$:
\begin{equation}\label{Eq_RVIal}
\nonumber
\begin{array}{ll}
 a^* \leftarrow  
 \arg\min_{a} 
 \left\{C(\underline{s})+\sum_{\underline{s}^{\prime} \in \underline{\mathcal{S}}} \operatorname{Pr}\left(\underline{s}^{\prime} \mid \underline{s}, a\right) V(\underline{s}^{\prime})\right\},
\\
 V(\underline{s}) \leftarrow\left\{C\left(\underline{s}\right)+
 \sum_{\underline{s}^{\prime} \in \underline{\mathcal{S}}}
 \operatorname{Pr}\left(\underline{s}^{\prime} \mid \underline{s}, a^*\right) V({\underline{s}^{\prime}})\right\}-V({\underline{s}_{\mathrm{ref}}}), 
\end{array}
\end{equation}
where $\underline{s}_{\mathrm{ref}} \in \underline{\mathcal{S}}$ is an arbitrarily chosen reference state. Once the iterative process above converges, the algorithm provides an optimal policy $\pi^*$ and the optimal value of the average AoII, which equals to $V({\underline{s}_{\mathrm{ref}}})$.
 \subsubsection{A Deep Q-Network (DQN) Policy to Solve \eqref{Eq_Prob_AoIIMDP}}
 Here the aim is to solve the MDP problem \eqref{Eq_Prob_AoIIMDP}. However, the main difficulty lies in the fact that the state space of the problem is an infinite set. Thus,  methods, e.g., RVI and linear programming \cite{Zakeri_CL}, which are applicable for problems with a \textit{finite} state space, cannot be utilized. 
 Nonetheless, problem \eqref{Eq_Prob_AoIIMDP} is an MDP problem and can be solved via online reinforcement learning algorithms.
 We adopt  a DQN \cite{Deep_Learning_Nature}  to solve problem \eqref{Eq_Prob_AoIIMDP}.
 A reader can refer to, e.g.,~{\cite[Alg.~1]{Deep_Learning_Nature}}, for more details of DQN. 
 Implementation details are given in the next section.

\section{ Numerical Results  }\label{Sec_NR}
Here, we provide simulation results to assess the performance of the derived policies. 
For the performance comparison, we also consider an AoI-optimal policy and a {``baseline policy"} which determines actions according to the following rule:  
{If $e(t) \ge \ct+\cs$, then ${a(t) = 2}$, i.e., the sampling and transmission action, else 
    $ a(t)=0$, i.e., the idle action}.
The sampling cost $c^{\mathrm{s}}$ and transmission cost $c^{\mathrm{t}}$ are fixed to $1$, and the value of $\bar{N}$ is set to $30$, unless specified otherwise. Furthermore, for the DQN policy,  we consider a fully-connected
deep neural network consisting of an input layer (${|z(t)|= N+3}$ neurons), $2$ hidden layers consisting of $64$ and $32$ neurons with \textit{ReLU} activation function, and an output layer (${|\mathcal{A}|=3}$ neurons); moreover, the number of steps per episode is $400$, the discount factor is $0.99$, the mini-batch size is $64$, the learning-rate is $0.0001$, and  the optimizer is \textit{RMSProp}.

\begin{figure}[t!]
\centering
\subfigure[ The average AoII vs. the self-transition probability for $\mu=0.5$  ] 
{
\includegraphics[width=0.3\textwidth]{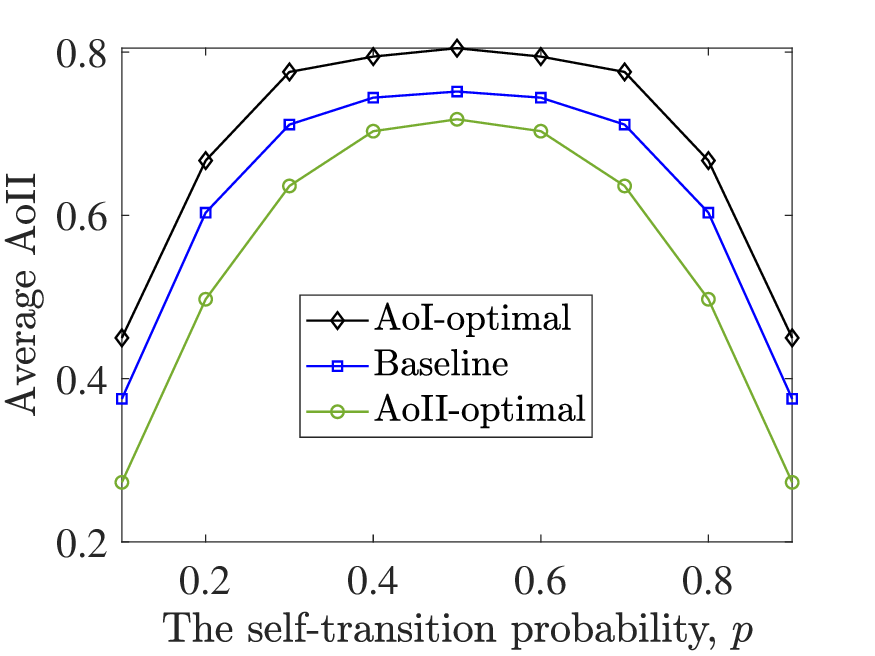}
\label{Fig_AoII_p}
}
\subfigure[ The average AoII vs. the energy arrival rate for $ p = 0.7$ ]{
\includegraphics[width=0.3\textwidth]{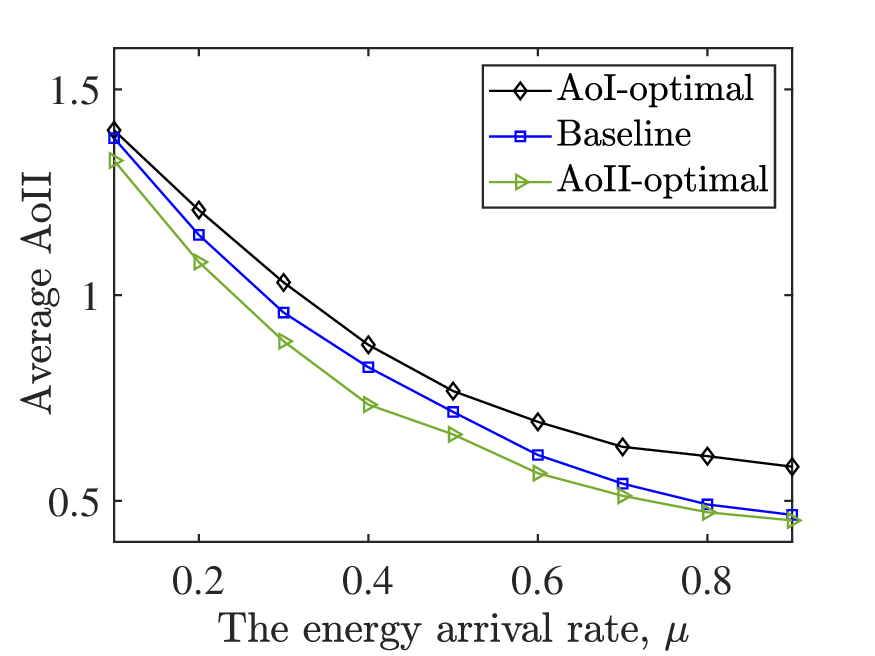}
\label{Fig_AoII_mu}
}
\caption{ The average AoII performance of the different policies, where $E=5$ and $q=1$ }
\label{Fig_AoII_allqcs}
\end{figure}

The average AoII performance of different policies is shown as a function of the self-transition probability of the source $p$ in Fig. \ref{Fig_AoII_p} and the energy arrival rate $\mu$ in Fig. \ref{Fig_AoII_mu}. Each policy is first optimized for the corresponding metric, and then its average AoII performance is calculated empirically. 
First, the figure shows that the AoII-optimal policy exhibits a significant improvement in performance compared to both the baseline policy and the AoI-optimal policy. 
This highlights the significance of considering the semantics of sampling and transmissions when optimizing the real-time tracking of a remote source, which is typically the primary goal in most status update systems.
Furthermore, it is observable that when the source undergoes rapid or gradual changes, its trackability increases owing to the predictability of the source state.  Besides, as expected, the performance for $p<0.5$ is identical to that of $p>0.5$. 

\begin{figure}[h!]
\centering
\subfigure[ The average AoII vs. the channel reliability $q$ for $\mu=0.5$  ] 
{
\includegraphics[width=0.31\textwidth]{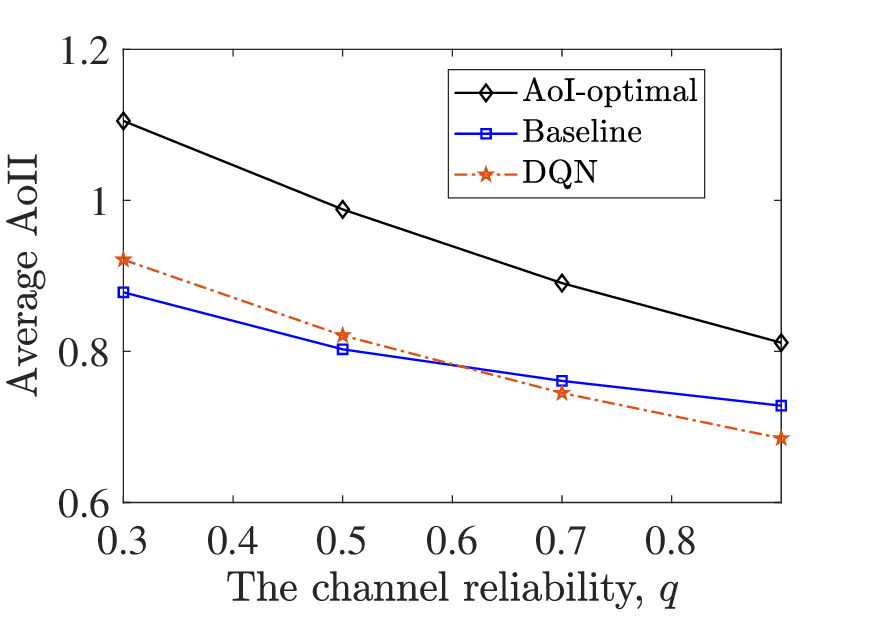}
\label{Fig_AoII_q}
}
\subfigure[ The average AoII vs. the sampling cost  for $ \mu = 0.7$ and $q=1$  ]{
\includegraphics[width=0.29\textwidth]{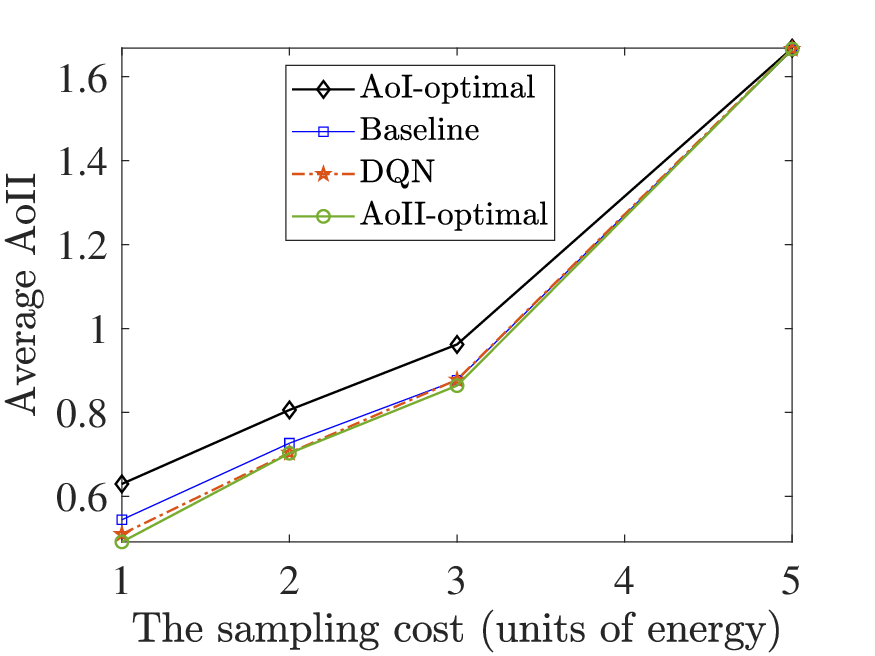}
\label{Fig_AoII_smc}
}
\caption{ The average AoII performance of the different policies, where $E=5$ and $p=0.7$ }
\label{Fig_AoII_qsmc}
\vspace{-2 em}
\end{figure}

 Figure \ref{Fig_AoII_q} demonstrates the average AoII as a function of the channel reliability $q$, where we use a DQN for the AoII optimization problem. (Reiterate that for the perfect channel, we obtained an AoII-optimal policy.)
 The figure shows that when the channel reliability is higher,  the DQN policy demonstrates a better performance since at a low reliable channel setup, finding optimal times of
sampling and transmission become more critical.
 
Finally, Figure \ref{Fig_AoII_smc} shows the average AoII with respect to the sampling cost. It reveals that the DQN policy coincides with the AoII-optimal policy. However, there exists a considerable performance gap between the AoI-optimal policy and the AoII-optimal policy when the sampling cost is small. 

\section{Conclusions}\label{Sec_Con}
We provided an AoII-optimal policy for real-time tracking in an energy harvesting system under sampling and transmission costs, where the sampling cost renders the source unobservable. 
To do so, we first formulated a stochastic optimization problem aimed at minimizing the  average AoII subject to the energy-causality constraint. 
We proposed a POMDP and its belief MDP formulation to tackle  the partial observability of the source, and we managed to effectively truncate the corresponding belief-state space and find an optimal policy when the channel is perfect. Moreover, for the general imperfect channel setup, a DQN policy is proposed.
Simulation experiments showed that the derived policies outperform the AoI-optimal policy and an opportunistic baseline policy almost in all circumstances. Additionally, they showed the source dynamic has a significant impact on the performance.

\bibliographystyle{ieeetr}
\bibliography{Bib_References/conf_short,
Bib_References/IEEEabrv,
Bib_References/Ref_2SRAoI}

\end{document}